\documentclass[aps,prb,twocolumn,showpacs]{revtex4-1}  
\usepackage{graphicx}  
\usepackage{epstopdf}
\usepackage{dcolumn}   
\usepackage{bm}        
\usepackage{amsmath}
\usepackage{amssymb}   
\usepackage{array}
\usepackage[caption=false]{subfig}
\usepackage[bookmarks=false,linkcolor=blue,urlcolor=blue,colorlinks,citecolor=blue]{hyperref}
\usepackage{dsfont}
\usepackage{appendix}

\makeatletter
\begin{document}

\title{Non-Abelian topological insulators from an array of quantum wires}

\author{Eran Sagi}

\author{Yuval Oreg}

\affiliation{Department of Condensed Matter Physics, Weizmann Institute of Science,
Rehovot, Israel 76100}

\date{\today}
\begin{abstract}
We suggest a construction of a large class of topological
states using an array of quantum wires. First, we show how to construct a Chern insulator using
an array of alternating wires that contain electrons and holes, correlated with an alternating
magnetic field. This
is supported by semi-classical arguments and a full quantum mechanical
treatment of an analogous tight-binding model. We then show how electron-electron
interactions can stabilize fractional Chern insulators (Abelian and non-Abelian). In particular, we construct a non-Abelian $\mathbb{Z}_{3}$ parafermion state. Our construction is
generalized to wires with alternating spin-orbit couplings,
which give rise to integer and fractional (Abelian and non-Abelian) topological
insulators. The states we construct are effectively two-dimensional, and
are therefore less sensitive to disorder than one-dimensional systems. The possibility of experimental realization of our construction
is addressed.
\end{abstract}
\maketitle

\section{Introduction}
 The integer quantum Hall effect (IQHE)~\cite{Klitzing1980} was discovered in two-dimensional (2D) systems subjected to a strong perpendicular
magnetic field.
The quantized conductance is a consequence of the emergence of a topological number ~\cite{Thouless1982}, known as the Chern number.
Haldane \cite{Haldane1988} showed that a Graphene-like material
which breaks time-reversal symmetry due to an alternating (zero average)
magnetic field may have a non-zero Chern number as well. These
types of materials, which have a non-zero Hall conductance with a zero total magnetic flux, are referred to as Chern insulators (CI).

The existence of edge modes \cite{Halperin1982} in the QHE can be understood
in various ways. In particular, one can understand the presence of edge
modes by studying the classical curved trajectories of electrons
in a magnetic field. In fact, it is possible to construct a semi-classical
theory for a specific set of Chern insulators as well. Consider a system consisting of electrons and holes (whose masses differ in sign). In the presence of a magnetic field, their classical trajectories are curved in opposite directions. If however, the electrons and the holes experience opposite magnetic fields, the trajectories will be curved in the same direction. One can imagine constructing a Chern-insulator by separating the plane into regions which contain only holes and only electrons. If the magnetic field is opposite in the two regions, the classical trajectories will be similar to those of electrons in a uniform magnetic field. This suggests that, upon quantization, this system should have a non-zero Chern number \cite{Note1}, despite the fact that the total magnetic flux vanishes.



Motivated by this semi-classical picture, we will study in this work
an effectively 2D system which consists of alternating wires that contain electrons
and holes. Approaching the 2D problem from the quasi-one-dimensional (Q1D) limit enables a full quantum-mechanical
analysis, and an analytic treatment of interaction effects using the
bosonization technique.
\begin{figure}[t]
\subfloat[\label{fig:system}]{\includegraphics[scale=0.25]{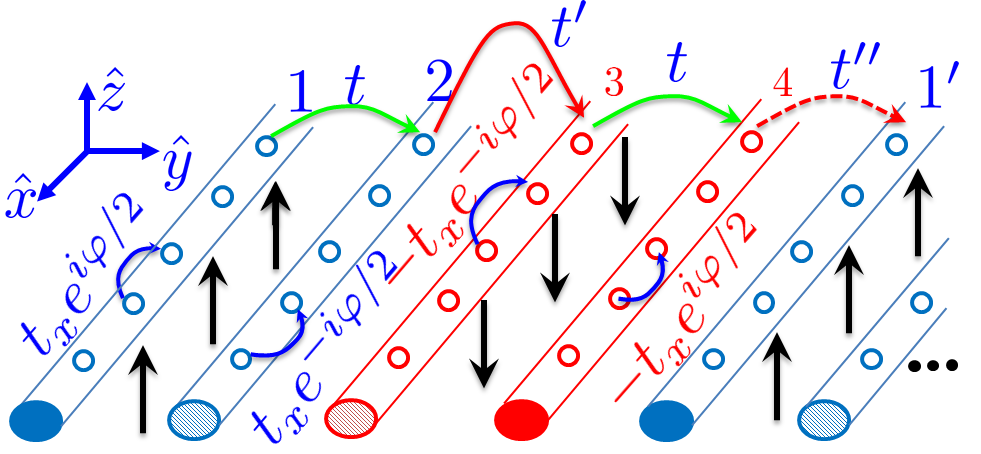}}
\subfloat[\label{fig:parabolas with no tunneling}]{\includegraphics[scale=0.25]{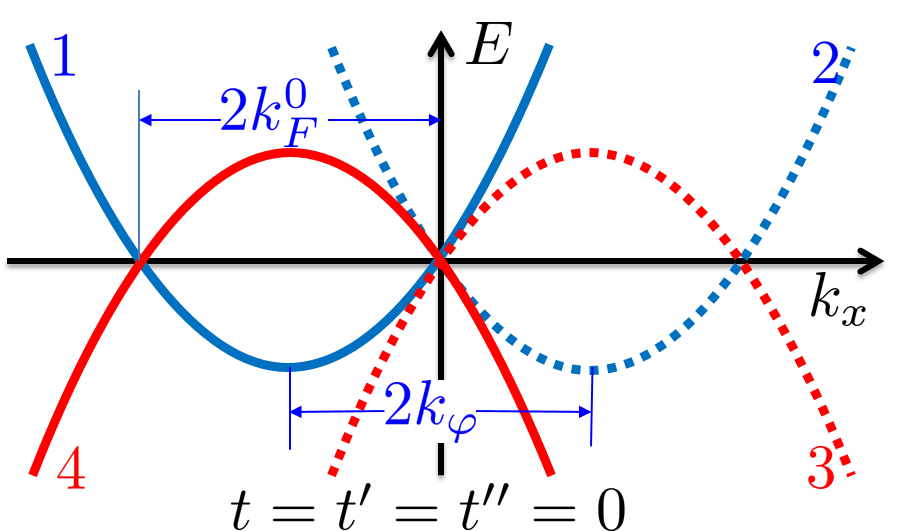}} \\
\subfloat[\label{fig:parabolas with tunneling}]{\includegraphics[scale=0.25]{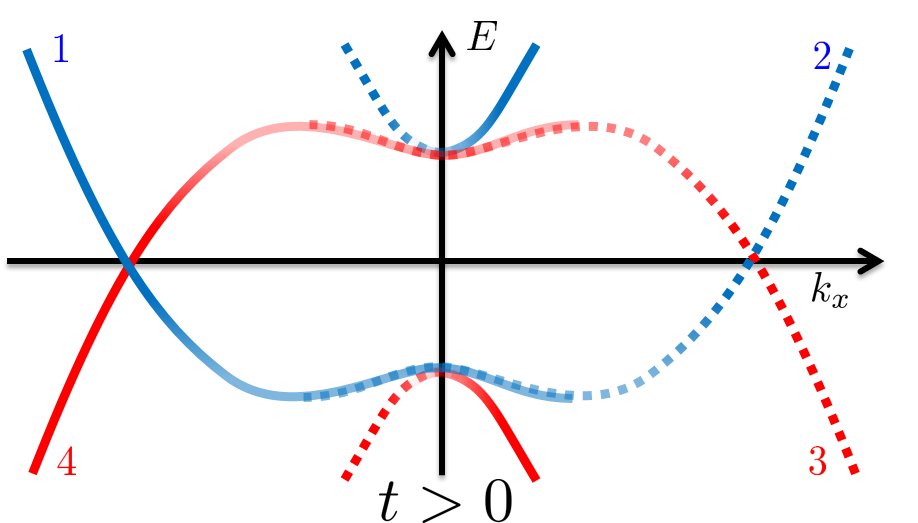}}
\subfloat[\label{fig:parabolas with tunneling2}]{\includegraphics[scale=0.25]{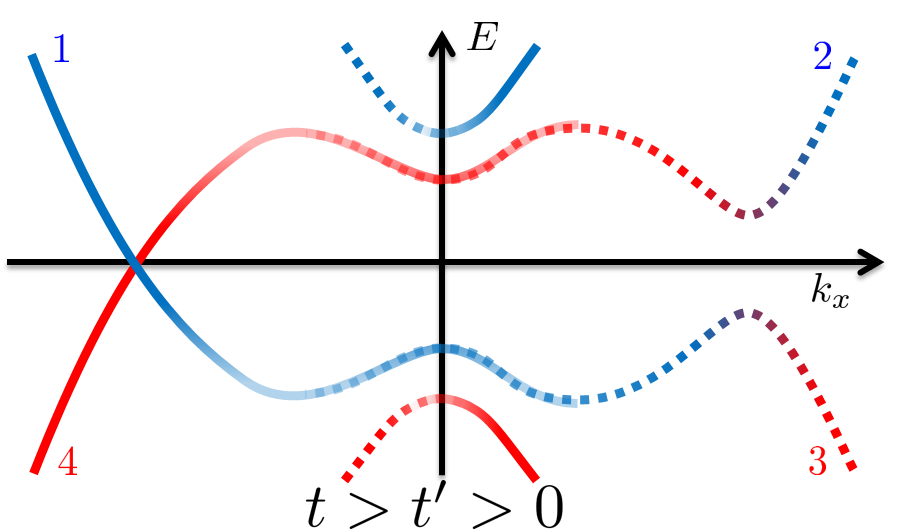}}
\caption{(Color Online) (a) Physical scheme of the Q1D model we study. Blue wires contain electrons, and red wires contain holes. The black arrows represent the magnetic field through the system. The circles represent the sites of the corresponding tight binding model~\cite{Supp}, and the tunneling amplitudes of the tight binding model are represented by colored arrows.  (b) The energy spectrum of the wires (as a function of $k_x$) near zero energy without tunneling between the wires ($t,t^\prime,t^{\prime\prime}=0$). The wires are tuned such that the four parabolas cross each other at zero energy, and the chemical potential is set to be zero. The spectra in blue, dashed blue, dashed red, and red correspond to wires 1, 2, 3, and 4 in a unit cell, respectively. 
(c) The energy spectrum when $t$ is switched on. A gap opens near $k_x=0$. (d) The spectrum when $t^\prime$ is switched on as well. This gives an additional gap at $k_x>0$. Free chiral modes are left on wires 1 and 4.
Finally, if one switches on $t^{\prime \prime}$, there are free chiral modes at the edge of the system, which suggests that there is a non-zero Chern number.
}
\end{figure}
Ref. \cite{Yakovenko1991,Sondhi2001} argues that it is possible to understand
the IQHE by considering a set of weakly coupled parallel wires.
First, we will show that 
one can use an array of wires to construct a CI as well (see Fig.~\ref{fig:system}).
We then introduce a tight-binding version of this model and obtain
a phase diagram, showing the Chern number as a function of the model
parameters.

Kane et al. \cite{Kane2002} generalized the wires approach to the
Abelian fractional quantum Hall effect (FQHE) using the bosonization technique.
We will generalize our construction to a fractional CI (FCI) as well.

To do so, we introduce composite particles.
This transformation maps the electrons and holes at 1/3 filling to composite particles at filling 1. 
The possibility of a FCI has recently been discussed quite extensively
in the literature \cite{parameswaran2013}. Numerical investigations \cite{Tang2011,Sun2011,Neupert2011,Sheng2011,Wang2012,Regnault2011} of lattice models with nearly flat bands presented strong evidence for FCI states. More general approaches, connecting the properties of the known FQHE states and analogous FCI states, were found \cite{Qi2011,parameswaran2012}.
Here we present an alternative analytic approach
to the subject, which may be applicable in experiments.

Teo \& Kane \cite{Teo2011} expand the approach of \cite{Kane2002}
to non-Abelian states. We will see
that our results can be generalized to the non-Abelian case as well,
and we will provide a detailed construction of a state similar
to the $\mathbb{Z}_{3}$ Read-Rezay state. This state supports Fibonacci anyons, which may be used for universal quantum computation \cite{Freedman2002,Freedman2002a}.

Using an analogy between a magnetic field and a spin-orbit coupling (in the $\hat z$ direction only), we will construct a topological insulator from an array of wires using an alternating spin-orbit coupling. It will then be straightforward to generalize the above model to a fractional topological insulator (FTI) \cite{Levin2009}. Other realizations of FTI states were discussed in Ref. \cite{Chen2012,Furukawa2014,Ghaemi2012,Neupert2011a,Repellin2014}.

We note that the wires approach was recently used by various papers \cite{Klinovaja2013a,Klinovaja2013c,Neupert2014} to discuss a variety of topological states.

\section{Wires construction of a CI}\label{sec:wires integer}
%
%
Motivated by the above semi-classical picture, we have designed the wires construction, shown in Fig.~\ref{fig:system}. In each unit cell there are four different wires. We tune the wires' chemical potentials such that wires 1 and 2 of each unit cell are near the bottom of the band, and wires 3 and 4 are near the top. Effectively, we have alternating pairs of wires that contain electrons and holes. A positive (negative) magnetic field is introduced between the pairs of electron (hole) wires. This is a Q1D version of the semi-classical picture we described above.

\textcolor{black}{For illustration and simplicity it is convenient to choose a gauge in which the vector potential $\mathbf{A}$ points at the $\hat x$ direction.
We can tune the wires' bands in such a way that all their crossing
points match in energy. In this case, the energy spectra are similar to those
depicted in Fig.~\ref{fig:parabolas with no tunneling}. We define $k_F^0$ as the Fermi momenta in the absence of an external magnetic field. $k{}_{\varphi} =\frac{eB a}{2\hbar c}$ is the shift of the parabolas due to the magnetic fields (see Fig. \ref{fig:parabolas with no tunneling}).}

If in addition, neighboring wires of the same type are weakly tunnel coupled (with an amplitude $t$),
a gap opens between parabolas 1 and 2, and parabolas 3 and 4. The spectrum
in this case is depicted in Fig.~\ref{fig:parabolas with tunneling}.

Introducing now a coupling between the electrons and holes \emph{inside} a unit cell ($t^\prime$), a gap will open at $k_x >0$, and we arrive at the spectrum
depicted in Fig.~\ref{fig:parabolas with tunneling2}.
If we now switch on small tunneling between different
unit cells ($t^{\prime \prime}$), the coupling between the edges decays exponentially with the sample width, and in the thermodynamic limit we expect to find gapless edge states.
The observation of gapless edge states
indicates that there is a non-zero Chern number.
To show this explicitly, we have constructed a 2D tight-binding
model, which is the lattice version of the above continuous model. The tight binding model enables an exact derivation of phase diagram, showing the Chern number as a function of the model parameters. For more details see the supplemental material \cite{Supp}, where the tight binding model is defined, and its phased diagram is derived, ensuring that the results of the above discussion are valid. We note that, by construction, our model has only a single chiral edge mode, leading to the fact that the model can only have Chern numbers $C=\pm1$. Generalization to larger Chern numbers is a possible interesting extension.
%


\section{FCI}
The wires construction invites us to add interactions and use bosonization techniques, similar to those used in \cite{Kane2002,Teo2011}. This allows us to generalize the above results to FCI states. In the presence of interactions, multi-electron processes may open a gap even if the Fermi point of the left movers is not equal to the Fermi points of the right movers~\cite{Kane2002,Teo2011,Oreg2013,Klinovaja2013}.

\begin{figure}[t]
\begin{center}
\subfloat[\label{fig:integer}]{\includegraphics[scale=0.3]{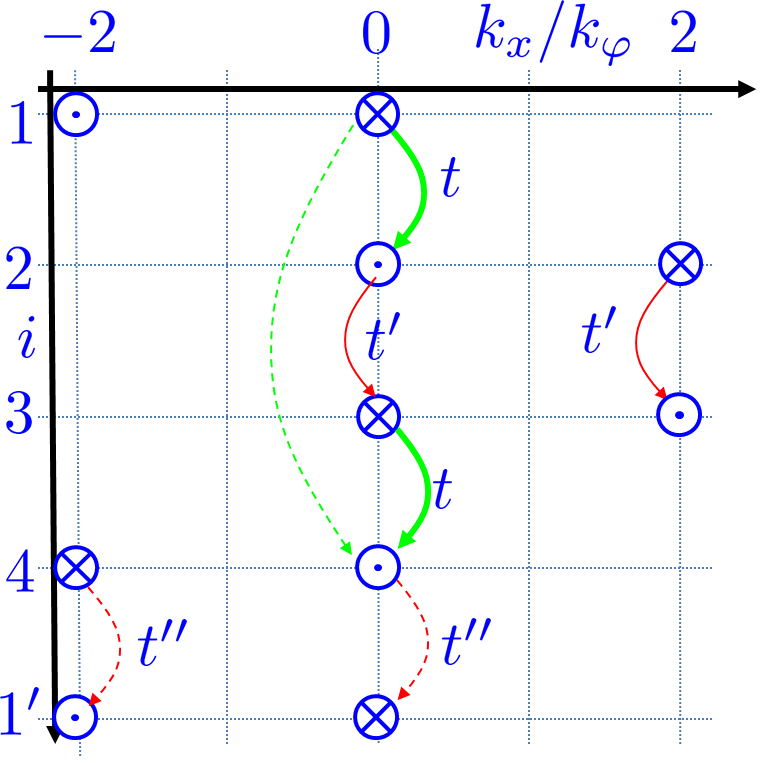}}
\subfloat[\label{fig:fractional}]{\includegraphics[scale=0.3]{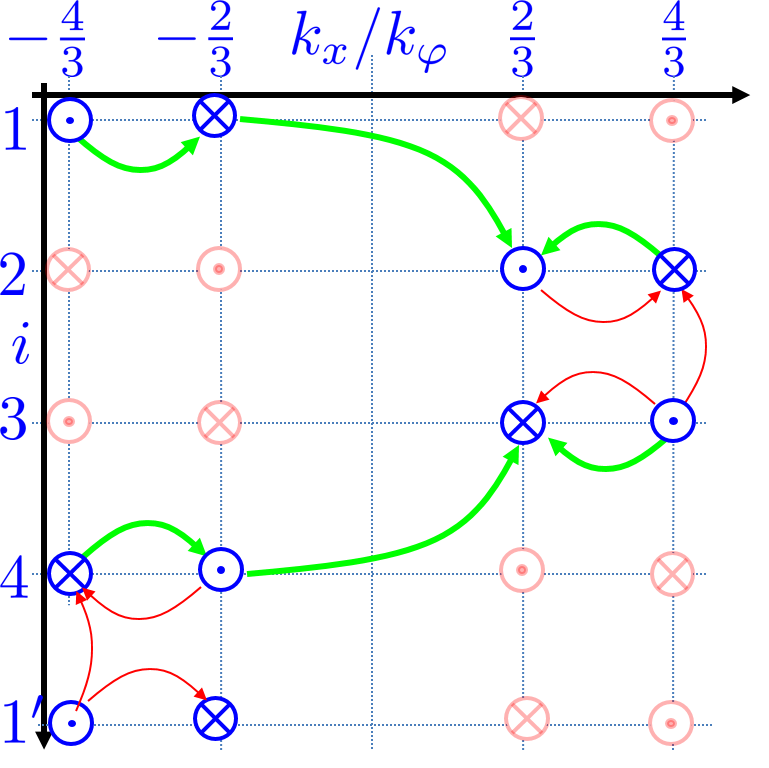}}
\end{center}
\caption{(Color Online) (a) A diagrammatic representation of the energy band
structure in the case $\nu \equiv k_F^0/k_\varphi=1$ (See Fig. \ref{fig:parabolas with no tunneling} for definitions of $k_\varphi$ and $k_F^0$). The $y$ axis shows the wire index inside
the unit cell, and the $x$ axis shows $k_x$ in units of $k_\varphi$. The symbol $\odot$ $(\otimes)$ represents $k_i^L$ ($k_i^R$).
 Colored arrows represent tunneling amplitudes between the wires. (b) The same diagram for a topological insulator
with $\nu=\frac{1}{3}$. Colored arrows now represent the multi-electron processes responsible for the creation of Laughlin-like states.
These complex processes in terms of the electrons ($\psi\sim e^{i\phi}$) for $\nu=1/3$ are mapped to simple tunneling processes in terms of the fermions $\tilde \psi \sim e^{i\eta}$. Thus, $\nu=1/3$ for $\psi$ is equivalent to $\nu=1$ for $\tilde \psi$.  In the presence of spin orbit coupling, spin up (blue) and spin down (light red) will experience opposite alternating effective magnetic fields.}
\end{figure}

To understand the required conditions for a gap opening due to multi-electron scattering processes, it is useful to present the spectra of Fig.~\ref{fig:parabolas with no tunneling} in an alternative way. Instead of plotting the entire spectrum of the wires together, we plot only the Fermi points as a function of the wire index. Soon, we will linearize the spectra around these points. A cross ($\bigotimes$) denotes the Fermi point of a right mover, and a dot ($\bigodot$) denotes the Fermi point of a left mover. Before analyzing the fractional case, it is useful to revisit the simple $\nu=1$ case. We will see that the main results of the tight binding model arise naturally in the bosonization framework. Fig.~\ref{fig:integer} shows the diagram that corresponds this case (Fig.~\ref{fig:parabolas with no tunneling}).  
Linearizing the spectrum around the Fermi-points of each wire, and using the standard bosonization procedure,
we define the two chiral bosonic fields, $\phi_{i}^R$ and $\phi_{i}^L$, for
each wire. In terms of these, the fermionic operators are
$\psi_{i}^R\propto e^{i\left(k_{i}^{R}x+\phi_{i}^R\right)},\psi_{i}^L\propto e^{i\left(k_{i}^{L}x+\phi_{i}^L\right)}.$

Without interactions, a momentum conserving single-electron tunneling between the wires (denoted in Fig.~\ref{fig:integer} by an arrow) is possible only when the left and right movers of adjacent wires are at the same point in $k$-space. The single-electron tunneling operators between adjacent wires (denoted in Fig.~\ref{fig:integer} by green, red, and dashed red arrows) are
\begin{align}
t \psi^{R\dagger}_{1(3)}\psi^L_{2(4)}+h.c. &\propto& t\cos\left(\phi_{1(3)}^R-\phi_{2(4)}^L \right),  \nonumber \\
t^\prime \psi^{R(L)\dagger}_{2}\psi^{L(R)}_{3} +h.c.  &\propto& t^\prime \cos\left(\phi_{2}^{R(L)}-\phi_{3}^{L(R)} \right), \nonumber \\
t^{\prime\prime} \psi^{R(L)\dagger}_{4}\psi^{L(R)}_{1^\prime} +h.c.  &\propto& t^{\prime \prime} \cos\left(\phi_{4}^{R(L)}-\phi_{1^\prime}^{L(R)} \right). \label{eq:tunneling}
\end{align}
We switch on the operators in the following way: first, we switch on a small $t\ll t_x$. Since this is a relevant operator, it gaps out the spectrum near $k_x=0$. Then, we switch on smaller electron-hole couplings $t^\prime,t^{\prime\prime}<t$.  The terms $\psi^{R\dagger}_{2}\psi^{L}_{3}$ and $\psi^{R\dagger}_{4}\psi^{L}_{1^\prime}$ gap out the rest of the spectrum, leaving a gapless edge mode. As we discussed before, this indicates that there is a non-zero Chern number. Note that the terms $\psi^{L\dagger}_{2}\psi^{R}_{3}$ and $\psi^{L\dagger}_{4}\psi^{R}_{1^\prime}$ contain fields which are conjugate to those already pinned by $t$. Strong quantum fluctuations are therefore expected to suppress these terms.

We now turn to generalize this to Laughlin-like FCI states, with a filling factor $\nu=k_F^0/k_\varphi=1/(2n+1)$, where $n$ is a non-negative integer. For example, the $k$-vector pattern of the wires with $\nu=1/3$ is shown in blue in Fig.~\ref{fig:fractional}. In this case, multi-electron processes are expected to gap out the system (except for the edges). To see this, it is enlightening to define new chiral fermion operators
\begin{align}
\tilde{\psi}_i^{R(L)} = (\psi_i^{R(L)})^{(n+1)} (\psi_i^{\dagger L(R)})^{n}  \propto  e^{i\left(q_{i}^{R(L)}x+\eta^{R(L)}_{i}\right)},
\end{align}
with
\begin{align}
\eta_{i}^{R(L)} =(n+1)\phi_i^{R(L)}-n \phi_i^{L(R)},
\end{align}
and $q_{i}^{R(L)}=(n+1)k_i^{R(L)}-nk_i^{L(R)}.$ A direct calculation of the commutation relations of the $\eta$-fields show that they have an additional factor of $2n\pi$ compared
the $\phi$-fields. This gives an extra (trivial) phase factor $e^{i2\pi n}$ in the anti-commutation relation of the $\tilde\psi$'s compared to the $\psi$'s,
insuring that the $\tilde \psi$'s are fermionic operators.  In addition, it can easily be checked that the resulting
structure of the $q$'s is identical to that of the $k$'s in the case of $\nu=1$ (Fig. \ref{fig:integer}), so that $\tilde{\psi}$ can be regarded as a fermionic field with $\nu=1$.
This procedure can therefore be interpreted as an attachment of $2n$ quantum fluxes to each electron
(cf. Jain's construction of composite fermions \cite{Jain2007}).

Repeating the analysis of the $\nu=1$ case,
we can now write single $\tilde \psi$ tunneling operators, identical to those found in Eq.~(\ref{eq:tunneling}) (replacing $\psi\rightarrow\tilde{\psi},\phi\rightarrow\eta$, with new tunneling amplitudes $\tilde{t},\tilde{t}^\prime$, and $\tilde{t}^{\prime\prime}$).
In terms of the original electrons, these operators describe the multi-electron processes shown in Fig.~\ref{fig:fractional}.
Note that when the interactions are strong enough, these operators become relevant \cite{Kane2002, Teo2011, Oreg2013}. From here, the process is identical to the integer case.
The gap due to the $\tilde\psi$ tunneling operators ensures that competing processes (for example, single electron tunneling between wires 2 and 3, or 4 and 1') are suppressed, as they contain fields that are conjugate to the fields pinned by $\tilde{t}$ (which is dominant by our construction).

The fact that the composite $\eta$-fields (and not the original $\phi$ fields) are pinned, leads to the various properties of these Laughlin-like states, like the fractional charge and statistics of the excitations, in analogy to the known FQHE states \cite{Kane2002,Teo2011}.

\section{Non-Abelian FCI}
As the discussion above shows, the wires construction allowed us to
create Abelian fractional Chern insulators. Ref.~\cite{Teo2011} constructed non-Abelian QHE states by enlarging the unit cell, and taking a non-uniform magnetic field inside each unit cell. By our construction, any non-Abelian state constructed by \cite{Teo2011} can be generalized to the CI case. To do so, one can take two unit cells from the construction in Ref.~\cite{Teo2011}, reverse the magnetic field of the second unit cell, and use holes instead of electrons. However, the lack of a total magnetic flux in our system enables a simpler construction of non-Abelian states, which don't have a direct analog in the QHE. We
now show that a slight modification of the procedure that enabled the construction of Laughlin-like states may lead to non-Abelian
states. We will focus here on a state similar to the $\mathbb{Z}_3$ Read-Rezay state.
Generalization to other non-Abelian states is possible.

To obtain a $\mathbb{Z}_3$ parafermion state, we take $\nu=1/3$,
and construct the $\tilde\psi $ operators. Let us start in the special point where $\tilde{t},\tilde{t}^\prime,$ and the coupling between $\tilde\psi^R_1$ and $\tilde\psi^L_4$ are tuned to have exactly the same value, denoted by $v$ (at the end, when the topological
nature of our construction will be revealed, this strict requirement
can be relaxed, as long as the bulk gap doesn't close). 
It can be shown (see supplemental material \cite{Supp} for more technical details) that under these assumptions our problem can be mapped to the $\beta^{2}=6\pi$ self-dual Sine-Gordon model,
which was studied in Ref.~\cite{Lecheminant2002,Mong2013}. Specifically, it was
shown that this model is mapped to a critical $\mathbb{Z}_{3}$
parafermionic field.

 We have established that any unit cell
has two counter propagating $\mathbb{Z}_3$ parafermionic fields (around $k=0$), and two counter propagating charge modes at $k_x=-2k_\varphi$.
 As earlier, we can in principle gap out the spectrum by switching on specific coupling terms between different unit cells, leaving
 eventually a Laughlin-like charge mode and a $\mathbb{Z}_3$
parafermion mode at the edge of the sample. However, in order to leave a chiral parafermion mode we need to consider quasiparticle tunneling terms, which are not allowed in the present construction \cite{Mong2013,Supp}. This technical problem is easily solved by adding an additional flavor quantum number to each wire, which allows one to effectively create a thin FQHE state in each unit cell (see supplemental material \cite{Supp}). This way, in addition to supporting gapless parafermion modes in each unit cell, our construction enables their coupling. We point out that the above construction is related to the bilayer QHE system presented by Ref. \cite{Vaezi2014}.
We note that while the non-Abelian part is the same as the non-Abelian part of the $\mathbb{Z}_3$ Read-Rezay state, the charge mode is different.

\section{Topological insulators from the wires approach}
\label{sec:topological insulators}
The entire analysis presented here can also be carried out for
\textcolor{black}{spinful} electrons if one introduces spin-orbit
interactions (in the $\hat z$ direction only). This can be done if an alternating electric field is
introduced instead of an alternating magnetic field. For example, the electric
field can be tuned in such a way that the spin-orbit coupling is
positive at wires 1 and 4, and negative at wires 2 and 3. Fig.~\ref{fig:fractional} shows the appropriate Fermi-momenta corresponding to $\nu=\frac{1}{3}$ (in blue for spin up
and light red for spin down).
If one considers only processes which conserve $S_z$, we get a simple construction for integer, Laughlin-like, and non-Abelian topological insulators \cite{Levin2009}, which are simply two copies of the FCI states discussed above (with opposite chiral modes for the different spin species). If we now introduce small time-reversal invariant terms which violate $S_z$ conservation (but do not close the gap), the helical modes remain protected. 

\section{Generalization}
The approach we present here can be extended \textcolor{black}{to hierarchical Abelian states, as well as other non-Abelian states} (such as a Moore-Read-like state).
One can also study the effects of proximity to a superconductor, which is expected to yield other non-Abelian states. A detailed further study of these constructions will be performed in the future.

{\section{Experimental realizations} The above theoretical construction may also be applicable in experiments with superlattices that realize the particle-hole structure we suggest. The alternating magnetic field can be generated, for example, using a snake-like wire \cite{Zeltzer}, as shown in Fig. \ref{fig:snake} (for more technical details see the supplemental material \cite{Supp}), or using an array of V-grooves \cite{Palevski}, as shown in Fig. \ref{fig:grooves}. We note that stripes with an alternating magnetic field can also be realized in cold atom systems \cite{Aidelsburger2011}.
\begin{figure}[t]
\begin{center}
\subfloat[\label{fig:snake}]{
{\includegraphics[scale=0.48]{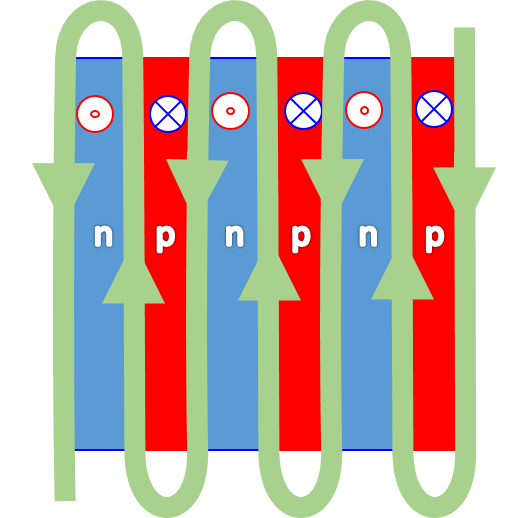}}}
\subfloat[\label{fig:grooves}]{{\includegraphics[scale=0.34]{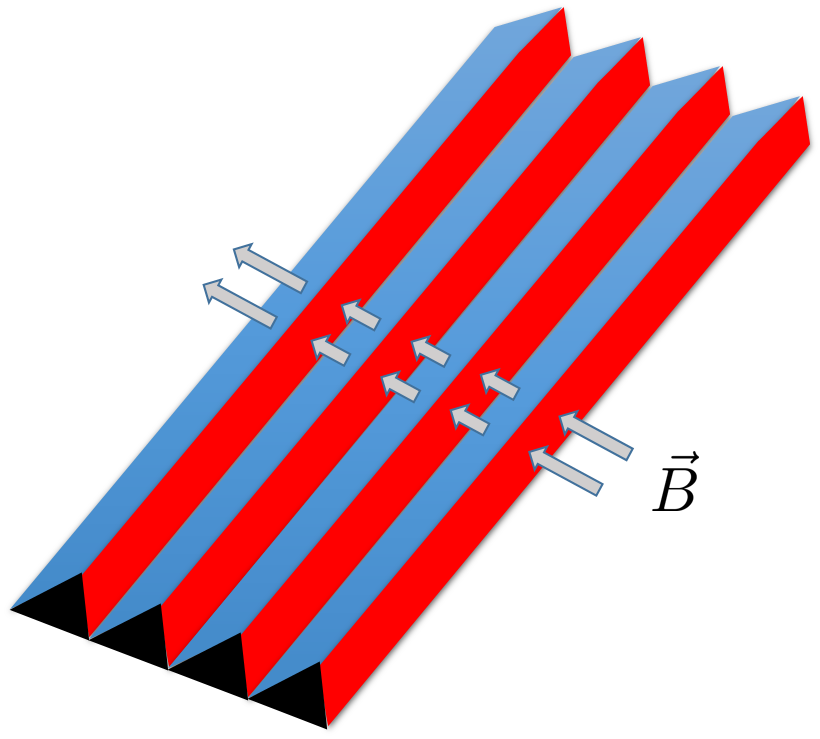}}}
\caption{(Color Online) (a) A possible experimental realization of our construction. Green lines represent current carrying wires, which produce the alternating magnetic field needed for our construction. This way, the magnetic field is positive (i.e. out of the page, denoted by $\bigodot$) in $n$-regions that contain electrons, and negative ($\bigotimes$) in $p$-regions that contain holes. (b) Another possible realization of our construction where now the 2D plane is replaced by many V-grooves connected in parallel. Again, blue regions contain electrons and red regions contain holes. The arrows represent the constant external magnetic field. In this geometry, electrons and holes experience opposite magnetic fields.}
\end{center}
\end{figure}
By coupling many (or maybe only a few) wires together we get an effective 2D system. Our construction lacks the disadvantages of fractional 1D
states \cite{Oreg2013,Klinovaja2013}, which are not topologically protected \cite{Fidkowski2011,Turner2011}, and the need to invoke proximity to a superconductor and a strong magnetic field simultaneously in 2D ~\cite{Lindner2012,Vaezi2013,Mong2013}. As long as the width of the edge modes is smaller than the sample width, it behaves practically as a topologically protected 2D system.

\begin{acknowledgments}
\emph{Acknowledgments}
We would like to thank Arbel Haim, Erez Berg, Eran Sela, Ady Stern, Netanel Lindner, Anna Kesselman, Jason Alicea, Gabriel Zeltzer and Alexander Palevski for useful discussions. We acknowledge the partial support of the Israeli Science Foundation (ISF), the Minerva foundation, the WIS-TAMU grant, and the European
Research Council under the European Community’s Seventh Framework Program
(FP7/2007-2013)/ERC Grant agreement No. 340210.
\end{acknowledgments}

\bibliographystyle{apsrev4-1}

\begin{thebibliography}{35}%
\makeatletter
\providecommand \@ifxundefined [1]{%
 \@ifx{#1\undefined}
}%
\providecommand \@ifnum [1]{%
 \ifnum #1\expandafter \@firstoftwo
 \else \expandafter \@secondoftwo
 \fi
}%
\providecommand \@ifx [1]{%
 \ifx #1\expandafter \@firstoftwo
 \else \expandafter \@secondoftwo
 \fi
}%
\providecommand \natexlab [1]{#1}%
\providecommand \enquote  [1]{``#1''}%
\providecommand \bibnamefont  [1]{#1}%
\providecommand \bibfnamefont [1]{#1}%
\providecommand \citenamefont [1]{#1}%
\providecommand \href@noop [0]{\@secondoftwo}%
\providecommand \href [0]{\begingroup \@sanitize@url \@href}%
\providecommand \@href[1]{\@@startlink{#1}\@@href}%
\providecommand \@@href[1]{\endgroup#1\@@endlink}%
\providecommand \@sanitize@url [0]{\catcode `\\12\catcode `\$12\catcode
  `\&12\catcode `\#12\catcode `\^12\catcode `\_12\catcode `\%12\relax}%
\providecommand \@@startlink[1]{}%
\providecommand \@@endlink[0]{}%
\providecommand \url  [0]{\begingroup\@sanitize@url \@url }%
\providecommand \@url [1]{\endgroup\@href {#1}{\urlprefix }}%
\providecommand \urlprefix  [0]{URL }%
\providecommand \Eprint [0]{\href }%
\providecommand \doibase [0]{http://dx.doi.org/}%
\providecommand \selectlanguage [0]{\@gobble}%
\providecommand \bibinfo  [0]{\@secondoftwo}%
\providecommand \bibfield  [0]{\@secondoftwo}%
\providecommand \translation [1]{[#1]}%
\providecommand \BibitemOpen [0]{}%
\providecommand \bibitemStop [0]{}%
\providecommand \bibitemNoStop [0]{.\EOS\space}%
\providecommand \EOS [0]{\spacefactor3000\relax}%
\providecommand \BibitemShut  [1]{\csname bibitem#1\endcsname}%
\let\auto@bib@innerbib\@empty
\bibitem [{\citenamefont {Klitzing}\ \emph {et~al.}(1980)\citenamefont
  {Klitzing}, \citenamefont {Dorda},\ and\ \citenamefont
  {Pepper}}]{Klitzing1980}%
  \BibitemOpen
  \bibfield  {author} {\bibinfo {author} {\bibfnamefont {K.}~\bibnamefont
  {Klitzing}}, \bibinfo {author} {\bibfnamefont {G.}~\bibnamefont {Dorda}}, \
  and\ \bibinfo {author} {\bibfnamefont {M.}~\bibnamefont {Pepper}},\ }\href
  {\doibase 10.1103/PhysRevLett.45.494} {\bibfield  {journal} {\bibinfo
  {journal} {Physical Review Letters}\ }\textbf {\bibinfo {volume} {45}},\
  \bibinfo {pages} {494} (\bibinfo {year} {1980})}\BibitemShut {NoStop}%
\bibitem [{\citenamefont {Thouless}\ \emph {et~al.}(1982)\citenamefont
  {Thouless}, \citenamefont {Kohmoto}, \citenamefont {Nightingale},\ and\
  \citenamefont {den Nijs}}]{Thouless1982}%
  \BibitemOpen
  \bibfield  {author} {\bibinfo {author} {\bibfnamefont {D.}~\bibnamefont
  {Thouless}}, \bibinfo {author} {\bibfnamefont {M.}~\bibnamefont {Kohmoto}},
  \bibinfo {author} {\bibfnamefont {M.}~\bibnamefont {Nightingale}}, \ and\
  \bibinfo {author} {\bibfnamefont {M.}~\bibnamefont {den Nijs}},\ }\href
  {\doibase 10.1103/PhysRevLett.49.405} {\bibfield  {journal} {\bibinfo
  {journal} {Physical Review Letters}\ }\textbf {\bibinfo {volume} {49}},\
  \bibinfo {pages} {405} (\bibinfo {year} {1982})}\BibitemShut {NoStop}%
\bibitem [{\citenamefont {Haldane}(1988)}]{Haldane1988}%
  \BibitemOpen
  \bibfield  {author} {\bibinfo {author} {\bibfnamefont {F.~D.~M.}\
  \bibnamefont {Haldane}},\ }\href {\doibase 10.1103/PhysRevLett.61.2015}
  {\bibfield  {journal} {\bibinfo  {journal} {Physical Review Letters}\
  }\textbf {\bibinfo {volume} {61}},\ \bibinfo {pages} {2015} (\bibinfo {year}
  {1988})}\BibitemShut {NoStop}%
\bibitem [{\citenamefont {Halperin}(1982)}]{Halperin1982}%
  \BibitemOpen
  \bibfield  {author} {\bibinfo {author} {\bibfnamefont {B.}~\bibnamefont
  {Halperin}},\ }\href {\doibase 10.1103/PhysRevB.25.2185} {\bibfield
  {journal} {\bibinfo  {journal} {Physical Review B}\ }\textbf {\bibinfo
  {volume} {25}},\ \bibinfo {pages} {2185} (\bibinfo {year}
  {1982})}\BibitemShut {NoStop}%
	\bibitem [{Note1()}]{Note1}%
  \BibitemOpen
  \bibinfo {note} {Note that since the transition between different regions is affected by the details of the model, we expect the Bohr-Sommerfeld quantization condition to provide bands which are not flat in general.}\BibitemShut {Stop}%
\bibitem [{\citenamefont {Yakovenko}(1991)}]{Yakovenko1991}%
  \BibitemOpen
  \bibfield  {author} {\bibinfo {author} {\bibfnamefont {V.}~\bibnamefont
  {Yakovenko}},\ }\href {\doibase 10.1103/PhysRevB.43.11353} {\bibfield
  {journal} {\bibinfo  {journal} {Physical Review B}\ }\textbf {\bibinfo
  {volume} {43}},\ \bibinfo {pages} {11353} (\bibinfo {year}
  {1991})}\BibitemShut {NoStop}%
\bibitem [{\citenamefont {Sondhi}\ and\ \citenamefont
  {Yang}(2001)}]{Sondhi2001}%
  \BibitemOpen
  \bibfield  {author} {\bibinfo {author} {\bibfnamefont {S.}~\bibnamefont
  {Sondhi}}\ and\ \bibinfo {author} {\bibfnamefont {K.}~\bibnamefont {Yang}},\
  }\href {\doibase 10.1103/PhysRevB.63.054430} {\bibfield  {journal} {\bibinfo
  {journal} {Physical Review B}\ }\textbf {\bibinfo {volume} {63}},\ \bibinfo
  {pages} {054430} (\bibinfo {year} {2001})}\BibitemShut {NoStop}%
\bibitem [{\citenamefont {Kane}\ \emph {et~al.}(2002)\citenamefont {Kane},
  \citenamefont {Mukhopadhyay},\ and\ \citenamefont {Lubensky}}]{Kane2002}%
  \BibitemOpen
  \bibfield  {author} {\bibinfo {author} {\bibfnamefont {C.}~\bibnamefont
  {Kane}}, \bibinfo {author} {\bibfnamefont {R.}~\bibnamefont {Mukhopadhyay}},
  \ and\ \bibinfo {author} {\bibfnamefont {T.}~\bibnamefont {Lubensky}},\
  }\href {\doibase 10.1103/PhysRevLett.88.036401} {\bibfield  {journal}
  {\bibinfo  {journal} {Physical Review Letters}\ }\textbf {\bibinfo {volume}
  {88}},\ \bibinfo {pages} {036401} (\bibinfo {year} {2002})}\BibitemShut
  {NoStop}%
\bibitem [{\citenamefont {Parameswaran}\ \emph {et~al.}(2013)\citenamefont
  {Parameswaran}, \citenamefont {Roy},\ and\ \citenamefont
  {Sondhi}}]{parameswaran2013}%
  \BibitemOpen
  \bibfield  {author} {\bibinfo {author} {\bibfnamefont {S.~A.}\ \bibnamefont
  {Parameswaran}}, \bibinfo {author} {\bibfnamefont {R.}~\bibnamefont {Roy}}, \
  and\ \bibinfo {author} {\bibfnamefont {S.~L.}\ \bibnamefont {Sondhi}},\
  }\href {\doibase 10.1016/j.crhy.2013.04.003} {\bibfield  {journal} {\bibinfo
  {journal} {Comptes Rendus Physique}\ }\textbf {\bibinfo {volume} {14}},\
  \bibinfo {pages} {816} (\bibinfo {year} {2013})}\BibitemShut {NoStop}%
\bibitem [{\citenamefont {Tang}\ \emph {et~al.}(2011)\citenamefont {Tang},
  \citenamefont {Mei},\ and\ \citenamefont {Wen}}]{Tang2011}%
  \BibitemOpen
  \bibfield  {author} {\bibinfo {author} {\bibfnamefont {E.}~\bibnamefont
  {Tang}}, \bibinfo {author} {\bibfnamefont {J.-W.}\ \bibnamefont {Mei}}, \
  and\ \bibinfo {author} {\bibfnamefont {X.-G.}\ \bibnamefont {Wen}},\ }\href
  {\doibase 10.1103/PhysRevLett.106.236802} {\bibfield  {journal} {\bibinfo
  {journal} {Physical Review Letters}\ }\textbf {\bibinfo {volume} {106}},\
  \bibinfo {pages} {236802} (\bibinfo {year} {2011})}\BibitemShut {NoStop}%
\bibitem [{\citenamefont {Sun}\ \emph {et~al.}(2011)\citenamefont {Sun},
  \citenamefont {Gu}, \citenamefont {Katsura},\ and\ \citenamefont {{Das
  Sarma}}}]{Sun2011}%
  \BibitemOpen
  \bibfield  {author} {\bibinfo {author} {\bibfnamefont {K.}~\bibnamefont
  {Sun}}, \bibinfo {author} {\bibfnamefont {Z.}~\bibnamefont {Gu}}, \bibinfo
  {author} {\bibfnamefont {H.}~\bibnamefont {Katsura}}, \ and\ \bibinfo
  {author} {\bibfnamefont {S.}~\bibnamefont {{Das Sarma}}},\ }\href {\doibase
  10.1103/PhysRevLett.106.236803} {\bibfield  {journal} {\bibinfo  {journal}
  {Physical Review Letters}\ }\textbf {\bibinfo {volume} {106}},\ \bibinfo
  {pages} {236803} (\bibinfo {year} {2011})}\BibitemShut {NoStop}%
\bibitem [{\citenamefont {Neupert}\ \emph {et~al.}(2011)\citenamefont
  {Neupert}, \citenamefont {Santos}, \citenamefont {Chamon},\ and\
  \citenamefont {Mudry}}]{Neupert2011}%
  \BibitemOpen
  \bibfield  {author} {\bibinfo {author} {\bibfnamefont {T.}~\bibnamefont
  {Neupert}}, \bibinfo {author} {\bibfnamefont {L.}~\bibnamefont {Santos}},
  \bibinfo {author} {\bibfnamefont {C.}~\bibnamefont {Chamon}}, \ and\ \bibinfo
  {author} {\bibfnamefont {C.}~\bibnamefont {Mudry}},\ }\href {\doibase
  10.1103/PhysRevLett.106.236804} {\bibfield  {journal} {\bibinfo  {journal}
  {Physical Review Letters}\ }\textbf {\bibinfo {volume} {106}},\ \bibinfo
  {pages} {236804} (\bibinfo {year} {2011})}\BibitemShut {NoStop}%
\bibitem [{\citenamefont {Sheng}\ \emph {et~al.}(2011)\citenamefont {Sheng},
  \citenamefont {Gu}, \citenamefont {Sun},\ and\ \citenamefont
  {Sheng}}]{Sheng2011}%
  \BibitemOpen
  \bibfield  {author} {\bibinfo {author} {\bibfnamefont {D.~N.}\ \bibnamefont
  {Sheng}}, \bibinfo {author} {\bibfnamefont {Z.-C.}\ \bibnamefont {Gu}},
  \bibinfo {author} {\bibfnamefont {K.}~\bibnamefont {Sun}}, \ and\ \bibinfo
  {author} {\bibfnamefont {L.}~\bibnamefont {Sheng}},\ }\href {\doibase
  10.1038/ncomms1380} {\bibfield  {journal} {\bibinfo  {journal} {Nature
  communications}\ }\textbf {\bibinfo {volume} {2}},\ \bibinfo {pages} {389}
  (\bibinfo {year} {2011})}\BibitemShut {NoStop}%
\bibitem [{\citenamefont {Wang}\ \emph {et~al.}(2012)\citenamefont {Wang},
  \citenamefont {Yao}, \citenamefont {Gu}, \citenamefont {Gong},\ and\
  \citenamefont {Sheng}}]{Wang2012}%
  \BibitemOpen
  \bibfield  {author} {\bibinfo {author} {\bibfnamefont {Y.-F.}\ \bibnamefont
  {Wang}}, \bibinfo {author} {\bibfnamefont {H.}~\bibnamefont {Yao}}, \bibinfo
  {author} {\bibfnamefont {Z.-C.}\ \bibnamefont {Gu}}, \bibinfo {author}
  {\bibfnamefont {C.-D.}\ \bibnamefont {Gong}}, \ and\ \bibinfo {author}
  {\bibfnamefont {D.~N.}\ \bibnamefont {Sheng}},\ }\href {\doibase
  10.1103/PhysRevLett.108.126805} {\bibfield  {journal} {\bibinfo  {journal}
  {Physical Review Letters}\ }\textbf {\bibinfo {volume} {108}},\ \bibinfo
  {pages} {126805} (\bibinfo {year} {2012})}\BibitemShut {NoStop}%
\bibitem [{\citenamefont {Regnault}\ and\ \citenamefont
  {Bernevig}(2011)}]{Regnault2011}%
  \BibitemOpen
  \bibfield  {author} {\bibinfo {author} {\bibfnamefont {N.}~\bibnamefont
  {Regnault}}\ and\ \bibinfo {author} {\bibfnamefont {B.~A.}\ \bibnamefont
  {Bernevig}},\ }\href {\doibase 10.1103/PhysRevX.1.021014} {\bibfield
  {journal} {\bibinfo  {journal} {Physical Review X}\ }\textbf {\bibinfo
  {volume} {1}},\ \bibinfo {pages} {021014} (\bibinfo {year}
  {2011})}\BibitemShut {NoStop}%
\bibitem [{\citenamefont {Qi}(2011)}]{Qi2011}%
  \BibitemOpen
  \bibfield  {author} {\bibinfo {author} {\bibfnamefont {X.-L.}\ \bibnamefont
  {Qi}},\ }\href {\doibase 10.1103/PhysRevLett.107.126803} {\bibfield
  {journal} {\bibinfo  {journal} {Physical Review Letters}\ }\textbf {\bibinfo
  {volume} {107}},\ \bibinfo {pages} {126803} (\bibinfo {year}
  {2011})}\BibitemShut {NoStop}%
\bibitem [{\citenamefont {Parameswaran}\ \emph {et~al.}(2012)\citenamefont
  {Parameswaran}, \citenamefont {Roy},\ and\ \citenamefont
  {Sondhi}}]{parameswaran2012}%
  \BibitemOpen
  \bibfield  {author} {\bibinfo {author} {\bibfnamefont {S.~A.}\ \bibnamefont
  {Parameswaran}}, \bibinfo {author} {\bibfnamefont {R.}~\bibnamefont {Roy}}, \
  and\ \bibinfo {author} {\bibfnamefont {S.~L.}\ \bibnamefont {Sondhi}},\
  }\href {\doibase 10.1103/PhysRevB.85.241308} {\bibfield  {journal} {\bibinfo
  {journal} {Physical Review B}\ }\textbf {\bibinfo {volume} {85}},\ \bibinfo
  {pages} {241308} (\bibinfo {year} {2012})}\BibitemShut {NoStop}%
\bibitem [{\citenamefont {Teo}\ and\ \citenamefont {Kane}(2011)}]{Teo2011}%
  \BibitemOpen
  \bibfield  {author} {\bibinfo {author} {\bibfnamefont {J.~C.~Y.}\
  \bibnamefont {Teo}}\ and\ \bibinfo {author} {\bibfnamefont {C.~L.}\
  \bibnamefont {Kane}},\ }\href {http://arxiv.org/abs/1111.2617}{\ (\bibinfo
  {year} {2011})},\ \Eprint {http://arxiv.org/abs/1111.2617}{arXiv:1111.2617}\BibitemShut {NoStop}%
\bibitem [{\citenamefont {Freedman}\ \emph
  {et~al.}(2002{\natexlab{a}})\citenamefont {Freedman}, \citenamefont
  {Larsen},\ and\ \citenamefont {Wang}}]{Freedman2002}%
  \BibitemOpen
  \bibfield  {author} {\bibinfo {author} {\bibfnamefont {M.~H.}\ \bibnamefont
  {Freedman}}, \bibinfo {author} {\bibfnamefont {M.~J.}\ \bibnamefont
  {Larsen}}, \ and\ \bibinfo {author} {\bibfnamefont {Z.}~\bibnamefont
  {Wang}},\ }\href {\doibase 10.1007/s002200200636} {\bibfield  {journal}
  {\bibinfo  {journal} {Communications in Mathematical Physics}\ }\textbf
  {\bibinfo {volume} {228}},\ \bibinfo {pages} {177} (\bibinfo {year}
  {2002}{\natexlab{a}})}\BibitemShut {NoStop}%
\bibitem [{\citenamefont {Freedman}\ \emph
  {et~al.}(2002{\natexlab{b}})\citenamefont {Freedman}, \citenamefont
  {Larsen},\ and\ \citenamefont {Wang}}]{Freedman2002a}%
  \BibitemOpen
  \bibfield  {author} {\bibinfo {author} {\bibfnamefont {M.~H.}\ \bibnamefont
  {Freedman}}, \bibinfo {author} {\bibfnamefont {M.}~\bibnamefont {Larsen}}, \
  and\ \bibinfo {author} {\bibfnamefont {Z.}~\bibnamefont {Wang}},\ }\href
  {\doibase 10.1007/s002200200645} {\bibfield  {journal} {\bibinfo  {journal}
  {Communications in Mathematical Physics}\ }\textbf {\bibinfo {volume}
  {227}},\ \bibinfo {pages} {605} (\bibinfo {year}
  {2002}{\natexlab{b}})}\BibitemShut {NoStop}%
  \bibitem [{\citenamefont {Levin}\ and\ \citenamefont
  {Stern}(2009)}]{Levin2009}%
  \BibitemOpen
  \bibfield  {author} {\bibinfo {author} {\bibfnamefont {M.}~\bibnamefont
  {Levin}}\ and\ \bibinfo {author} {\bibfnamefont {A.}~\bibnamefont {Stern}},\
  }\href {\doibase 10.1103/PhysRevLett.103.196803} {\bibfield  {journal}
  {\bibinfo  {journal} {Physical Review Letters}\ }\textbf {\bibinfo {volume}
  {103}},\ \bibinfo {pages} {196803} (\bibinfo {year} {2009})}\BibitemShut
  {NoStop}%
\bibitem [{\citenamefont {Chen}\ and\ \citenamefont {Yang}(2012)}]{Chen2012}%
  \BibitemOpen
  \bibfield  {author} {\bibinfo {author} {\bibfnamefont {H.}~\bibnamefont
  {Chen}}\ and\ \bibinfo {author} {\bibfnamefont {K.}~\bibnamefont {Yang}},\
  }\href {\doibase 10.1103/PhysRevB.85.195113} {\bibfield  {journal} {\bibinfo
  {journal} {Physical Review B}\ }\textbf {\bibinfo {volume} {85}},\ \bibinfo
  {pages} {195113} (\bibinfo {year} {2012})}\BibitemShut {NoStop}%
\bibitem [{\citenamefont {Furukawa}\ and\ \citenamefont
  {Ueda}(2014)}]{Furukawa2014}%
  \BibitemOpen
  \bibfield  {author} {\bibinfo {author} {\bibfnamefont {S.}~\bibnamefont
  {Furukawa}}\ and\ \bibinfo {author} {\bibfnamefont {M.}~\bibnamefont
  {Ueda}},\ }\href {http://xxx.tau.ac.il/abs/1402.6860} {\  (\bibinfo {year}
  {2014})},\ \Eprint {http://arxiv.org/abs/1402.6860} {arXiv:1402.6860}
  \BibitemShut {NoStop}%
\bibitem [{\citenamefont {Ghaemi}\ \emph {et~al.}(2012)\citenamefont {Ghaemi},
  \citenamefont {Cayssol}, \citenamefont {Sheng},\ and\ \citenamefont
  {Vishwanath}}]{Ghaemi2012}%
  \BibitemOpen
  \bibfield  {author} {\bibinfo {author} {\bibfnamefont {P.}~\bibnamefont
  {Ghaemi}}, \bibinfo {author} {\bibfnamefont {J.}~\bibnamefont {Cayssol}},
  \bibinfo {author} {\bibfnamefont {D.~N.}\ \bibnamefont {Sheng}}, \ and\
  \bibinfo {author} {\bibfnamefont {A.}~\bibnamefont {Vishwanath}},\ }\href
  {\doibase 10.1103/PhysRevLett.108.266801} {\bibfield  {journal} {\bibinfo
  {journal} {Physical Review Letters}\ }\textbf {\bibinfo {volume} {108}},\
  \bibinfo {pages} {266801} (\bibinfo {year} {2012})}\BibitemShut {NoStop}%
\bibitem [{\citenamefont {Neupert}\ \emph
  {et~al.}(2011{\natexlab{b}})\citenamefont {Neupert}, \citenamefont {Santos},
  \citenamefont {Ryu}, \citenamefont {Chamon},\ and\ \citenamefont
  {Mudry}}]{Neupert2011a}%
  \BibitemOpen
  \bibfield  {author} {\bibinfo {author} {\bibfnamefont {T.}~\bibnamefont
  {Neupert}}, \bibinfo {author} {\bibfnamefont {L.}~\bibnamefont {Santos}},
  \bibinfo {author} {\bibfnamefont {S.}~\bibnamefont {Ryu}}, \bibinfo {author}
  {\bibfnamefont {C.}~\bibnamefont {Chamon}}, \ and\ \bibinfo {author}
  {\bibfnamefont {C.}~\bibnamefont {Mudry}},\ }\href {\doibase
  10.1103/PhysRevB.84.165107} {\bibfield  {journal} {\bibinfo  {journal}
  {Physical Review B}\ }\textbf {\bibinfo {volume} {84}},\ \bibinfo {pages}
  {165107} (\bibinfo {year} {2011}{\natexlab{b}})}\BibitemShut {NoStop}%
\bibitem [{\citenamefont {Repellin}\ \emph {et~al.}(2014)\citenamefont
  {Repellin}, \citenamefont {Bernevig},\ and\ \citenamefont
  {Regnault}}]{Repellin2014}%
  \BibitemOpen
  \bibfield  {author} {\bibinfo {author} {\bibfnamefont {C.}~\bibnamefont
  {Repellin}}, \bibinfo {author} {\bibfnamefont {B.~A.}\ \bibnamefont
  {Bernevig}}, \ and\ \bibinfo {author} {\bibfnamefont {N.}~\bibnamefont
  {Regnault}},\ }\href {http://arxiv.org/abs/1402.2652} {\  (\bibinfo {year}
  {2014})},\ \Eprint {http://arxiv.org/abs/1402.2652} {arXiv:1402.2652}
  \BibitemShut {NoStop}%
\bibitem [{\citenamefont {Oreg}\ \emph {et~al.}(2013)\citenamefont {Oreg},
  \citenamefont {Sela},\ and\ \citenamefont {Stern}}]{Oreg2013}%
  \BibitemOpen
  \bibfield  {author} {\bibinfo {author} {\bibfnamefont {Y.}~\bibnamefont
  {Oreg}}, \bibinfo {author} {\bibfnamefont {E.}~\bibnamefont {Sela}}, \ and\
  \bibinfo {author} {\bibfnamefont {A.}~\bibnamefont {Stern}},\ }\href
  {http://arxiv.org/abs/1301.7335} {\  (\bibinfo {year} {2013})},\ \Eprint
  {http://arxiv.org/abs/1301.7335} {arXiv:1301.7335}\BibitemShut {NoStop}%
\bibitem [{\citenamefont {Klinovaja}\ and\ \citenamefont
  {Loss}(2013{\natexlab{a}})}]{Klinovaja2013}%
  \BibitemOpen
  \bibfield  {author} {\bibinfo {author} {\bibfnamefont {J.}~\bibnamefont
  {Klinovaja}}\ and\ \bibinfo {author} {\bibfnamefont {D.}~\bibnamefont
  {Loss}},\ }\href {http://arxiv.org/abs/1311.3259} {\  (\bibinfo {year}
  {2013}{\natexlab{a}})},\ \Eprint {http://arxiv.org/abs/1311.3259}
  {arXiv:1311.3259}\BibitemShut {NoStop}%
\bibitem [{\citenamefont {Lindner}\ \emph {et~al.}(2012)\citenamefont
  {Lindner}, \citenamefont {Berg}, \citenamefont {Refael},\ and\ \citenamefont
  {Stern}}]{Lindner2012}%
  \BibitemOpen
  \bibfield  {author} {\bibinfo {author} {\bibfnamefont {N.~H.}\ \bibnamefont
  {Lindner}}, \bibinfo {author} {\bibfnamefont {E.}~\bibnamefont {Berg}},
  \bibinfo {author} {\bibfnamefont {G.}~\bibnamefont {Refael}}, \ and\ \bibinfo
  {author} {\bibfnamefont {A.}~\bibnamefont {Stern}},\ }\href {\doibase
  10.1103/PhysRevX.2.041002} {\bibfield  {journal} {\bibinfo  {journal}
  {Physical Review X}\ }\textbf {\bibinfo {volume} {2}},\ \bibinfo {pages}
  {041002} (\bibinfo {year} {2012})}\BibitemShut {NoStop}%
\bibitem [{\citenamefont {Vaezi}(2013)}]{Vaezi2013}%
  \BibitemOpen
  \bibfield  {author} {\bibinfo {author} {\bibfnamefont {A.}~\bibnamefont
  {Vaezi}},\ }\href {http://arxiv.org/abs/1307.8069} {\  (\bibinfo {year}
  {2013})},\ \Eprint {http://arxiv.org/abs/1307.8069} {arXiv:1307.8069}\BibitemShut {NoStop}%
\bibitem [{\citenamefont {Mong}\ \emph {et~al.}(2013)\citenamefont {Mong},
  \citenamefont {Clarke}, \citenamefont {Alicea}, \citenamefont {Lindner},
  \citenamefont {Fendley}, \citenamefont {Nayak}, \citenamefont {Oreg},
  \citenamefont {Stern}, \citenamefont {Berg}, \citenamefont {Shtengel},\ and\
  \citenamefont {Fisher}}]{Mong2013}%
  \BibitemOpen
  \bibfield  {author} {\bibinfo {author} {\bibfnamefont {R.~S.~K.}\
  \bibnamefont {Mong}}, \bibinfo {author} {\bibfnamefont {D.~J.}\ \bibnamefont
  {Clarke}}, \bibinfo {author} {\bibfnamefont {J.}~\bibnamefont {Alicea}},
  \bibinfo {author} {\bibfnamefont {N.~H.}\ \bibnamefont {Lindner}}, \bibinfo
  {author} {\bibfnamefont {P.}~\bibnamefont {Fendley}}, \bibinfo {author}
  {\bibfnamefont {C.}~\bibnamefont {Nayak}}, \bibinfo {author} {\bibfnamefont
  {Y.}~\bibnamefont {Oreg}}, \bibinfo {author} {\bibfnamefont {A.}~\bibnamefont
  {Stern}}, \bibinfo {author} {\bibfnamefont {E.}~\bibnamefont {Berg}},
  \bibinfo {author} {\bibfnamefont {K.}~\bibnamefont {Shtengel}}, \ and\
  \bibinfo {author} {\bibfnamefont {M.~P.~A.}\ \bibnamefont {Fisher}},\ }\href
  {http://arxiv.org/abs/1307.4403} {\  (\bibinfo {year} {2013})},\ \Eprint
  {http://arxiv.org/abs/1307.4403} {arXiv:1307.4403}\BibitemShut{NoStop}%
\bibitem [{\citenamefont {Klinovaja}\ and\ \citenamefont
  {Loss}(2013{\natexlab{b}})}]{Klinovaja2013a}%
  \BibitemOpen
  \bibfield  {author} {\bibinfo {author} {\bibfnamefont {J.}~\bibnamefont
  {Klinovaja}}\ and\ \bibinfo {author} {\bibfnamefont {D.}~\bibnamefont
  {Loss}},\ }\href {http://arxiv.org/abs/1305.1569} {\  (\bibinfo {year}
  {2013}{\natexlab{b}})},\ \Eprint {http://arxiv.org/abs/1305.1569}
  {arXiv:1305.1569}\BibitemShut{NoStop}%
\bibitem [{\citenamefont {Klinovaja}\ and\ \citenamefont
  {Loss}(2013{\natexlab{c}})}]{Klinovaja2013c}%
  \BibitemOpen
  \bibfield  {author} {\bibinfo {author} {\bibfnamefont {J.}~\bibnamefont
  {Klinovaja}}\ and\ \bibinfo {author} {\bibfnamefont {D.}~\bibnamefont
  {Loss}},\ }\href {\doibase 10.1103/PhysRevLett.111.196401} {\bibfield
  {journal} {\bibinfo  {journal} {Physical Review Letters}\ }\textbf {\bibinfo
  {volume} {111}},\ \bibinfo {pages} {196401} (\bibinfo {year}
  {2013}{\natexlab{c}})}\BibitemShut {NoStop}
\bibitem [{\citenamefont {Neupert}\ \emph {et~al.}(2014)\citenamefont
  {Neupert}, \citenamefont {Chamon}, \citenamefont {Mudry},\ and\ \citenamefont
  {Thomale}}]{Neupert2014}%
  \BibitemOpen
  \bibfield  {author} {\bibinfo {author} {\bibfnamefont {T.}~\bibnamefont
  {Neupert}}, \bibinfo {author} {\bibfnamefont {C.}~\bibnamefont {Chamon}},
  \bibinfo {author} {\bibfnamefont {C.}~\bibnamefont {Mudry}}, \ and\ \bibinfo
  {author} {\bibfnamefont {R.}~\bibnamefont {Thomale}},\ }\href
  {http://arxiv.org/abs/1403.0953} {\  (\bibinfo {year} {2014})},\ \Eprint
  {http://arxiv.org/abs/1403.0953} {arXiv:1403.0953}\BibitemShut{NoStop}%
\bibitem [{Sup()}]{Supp}%
  \BibitemOpen\href@noop{}{\bibinfo{title}{See supplemental material.}}\BibitemShut{Stop}%

\bibitem [{\citenamefont {Jain}(2007)}]{Jain2007}%
  \BibitemOpen
  \bibfield  {author} {\bibinfo {author} {\bibfnamefont {J.~K.}\ \bibnamefont
  {Jain}},\ }\href
  {http://www.amazon.com/Composite-Fermions-Jainendra-K-Jain/dp/0521862329}
  {\emph {\bibinfo {title} {{Composite Fermions}}}}\ (\bibinfo  {publisher}
  {Cambridge University Press},\ \bibinfo {year} {2007})\ p.\ \bibinfo {pages}
  {560}\BibitemShut {NoStop}%
\bibitem [{\citenamefont {Lecheminant}\ \emph {et~al.}(2002)\citenamefont
  {Lecheminant}, \citenamefont {Gogolin},\ and\ \citenamefont
  {Nersesyan}}]{Lecheminant2002}%
  \BibitemOpen
  \bibfield  {author} {\bibinfo {author} {\bibfnamefont {P.}~\bibnamefont
  {Lecheminant}}, \bibinfo {author} {\bibfnamefont {A.~O.}\ \bibnamefont
  {Gogolin}}, \ and\ \bibinfo {author} {\bibfnamefont {A.~A.}\ \bibnamefont
  {Nersesyan}},\ }\href
  {http://www.sciencedirect.com/science/article/pii/S0550321302004741}
  {\bibfield  {journal} {\bibinfo  {journal} {Nuclear Physics B}\ }\textbf
  {\bibinfo {volume} {639}},\ \bibinfo {pages} {502} (\bibinfo {year}
  {2002})}\BibitemShut {NoStop}%
\bibitem [{\citenamefont {Vaezi}\ and\ \citenamefont
  {Barkeshli}(2014)}]{Vaezi2014}%
  \BibitemOpen
  \bibfield  {author} {\bibinfo {author} {\bibfnamefont {A.}~\bibnamefont
  {Vaezi}}\ and\ \bibinfo {author} {\bibfnamefont {M.}~\bibnamefont
  {Barkeshli}},\ }\href
  {http://arxiv.org/abs/1403.3383} {\  (\bibinfo {year} {2014})},\ \Eprint
  {http://arxiv.org/abs/1403.3383} {arXiv:1403.3383}\BibitemShut{NoStop}%
\bibitem [{Zel()}]{Zeltzer}%
  \BibitemOpen
  \href@noop {} {{\bibinfo {title} {{G. Zeltzer (private
  communication).}}}\ }\BibitemShut {Stop}%
	\bibitem [{Pal()}]{Palevski}%
  \BibitemOpen
  \href@noop {} {{\bibinfo {title} {{A. Palevski (private
  communication).}}}\ }\BibitemShut {Stop}%
\bibitem [{\citenamefont {Aidelsburger}\ \emph {et~al.}(2011)\citenamefont
  {Aidelsburger}, \citenamefont {Atala}, \citenamefont {Nascimb\`{e}ne},
  \citenamefont {Trotzky}, \citenamefont {Chen},\ and\ \citenamefont
  {Bloch}}]{Aidelsburger2011}%
  \BibitemOpen
  \bibfield  {author} {\bibinfo {author} {\bibfnamefont {M.}~\bibnamefont
  {Aidelsburger}}, \bibinfo {author} {\bibfnamefont {M.}~\bibnamefont {Atala}},
  \bibinfo {author} {\bibfnamefont {S.}~\bibnamefont {Nascimb\`{e}ne}},
  \bibinfo {author} {\bibfnamefont {S.}~\bibnamefont {Trotzky}}, \bibinfo
  {author} {\bibfnamefont {Y.-A.}\ \bibnamefont {Chen}}, \ and\ \bibinfo
  {author} {\bibfnamefont {I.}~\bibnamefont {Bloch}},\ }\href {\doibase
  10.1103/PhysRevLett.107.255301} {\bibfield  {journal} {\bibinfo  {journal}
  {Physical Review Letters}\ }\textbf {\bibinfo {volume} {107}},\ \bibinfo
  {pages} {255301} (\bibinfo {year} {2011})}\BibitemShut {NoStop}%
\bibitem [{\citenamefont {Fidkowski}\ and\ \citenamefont
  {Kitaev}(2011)}]{Fidkowski2011}%
  \BibitemOpen
  \bibfield  {author} {\bibinfo {author} {\bibfnamefont {L.}~\bibnamefont
  {Fidkowski}}\ and\ \bibinfo {author} {\bibfnamefont {A.}~\bibnamefont
  {Kitaev}},\ }\href {\doibase 10.1103/PhysRevB.83.075103} {\bibfield
  {journal} {\bibinfo  {journal} {Physical Review B}\ }\textbf {\bibinfo
  {volume} {83}},\ \bibinfo {pages} {075103} (\bibinfo {year}
  {2011})}\BibitemShut {NoStop}%
\bibitem [{\citenamefont {Turner}\ \emph {et~al.}(2011)\citenamefont {Turner},
  \citenamefont {Pollmann},\ and\ \citenamefont {Berg}}]{Turner2011}%
  \BibitemOpen
  \bibfield  {author} {\bibinfo {author} {\bibfnamefont {A.~M.}\ \bibnamefont
  {Turner}}, \bibinfo {author} {\bibfnamefont {F.}~\bibnamefont {Pollmann}}, \
  and\ \bibinfo {author} {\bibfnamefont {E.}~\bibnamefont {Berg}},\ }\href
  {\doibase 10.1103/PhysRevB.83.075102} {\bibfield  {journal} {\bibinfo
  {journal} {Physical Review B}\ }\textbf {\bibinfo {volume} {83}},\ \bibinfo
  {pages} {075102} (\bibinfo {year} {2011})}\BibitemShut {NoStop}%
\end{thebibliography}

\onecolumngrid
\newpage

\section*{\large{Supplemental Material}}
\onecolumngrid

\maketitle
This supplemental material introduces the tight binding
version of the wires construction. A detailed derivation of the phase
diagram is presented. In addition, we show explicitly how our construction of a non-Abelian state is mapped to the self-dual Sine-Gordon model, and provide technical details about
the experimental realization suggested at the concluding part of the main text.

\section*{Tight binding model}
To show explicitly that the model is topologically non-trivial, we have constructed a 2D tight-binding
model, which is the lattice version of the wires construction defined in the main text.

The unit cell of the
model is larger than the basic unit cell of the underlying square lattice: each unit cell contains 4 points, corresponding to the 4 different
wires in Fig. 1a of the main text. Each point is characterized by $(x,y)$,
the location of the unit cell (measured in units of the lattice spacing
$a$), and an index $n=1,2,3,4$ which labels the sites inside the unit cell.

We introduce the nearest neighbor tunneling amplitudes: $t,t^{\prime},t^{\prime\prime},\pm t_{x}e^{\pm ik_{\varphi}a}$,
described in Fig. 1a of the main text.
The phase $e^{\pm ik_{\varphi}a}$, with $k{}_{\varphi} a=\frac{eB a^2}{2\hbar c}$, is the Peierls phase associated with the alternating external magnetic
field. It is related to the flux in a basic unit cell: $k{}_{\varphi} a=\frac{\varphi}{2}$, where $\varphi=2\pi \frac{\Phi}{\Phi_0}$.
 In addition, we introduce a mass term
(not related to the physical mass of the particle), which adds a constant
energy $m$ ($-m$) for electrons (holes). The sign of $t_{x}$ is equal to the sign
of $m$. 
Similar to Haldane's original model, $m$ and $\varphi$ tune the system into and out of the
topological phases.
For simplicity, we restrict ourselves to the case $t^{\prime\prime}=t^{\prime}$.


We define the annihilation operators, $c_{n}(x,y)$, in terms of which the tight binding Hamiltonian described in the main text is a sum
of the following terms:
\begin{equation}
H_{m}=m\sum_{x,y}\left[\sum_{n=1}^{2}c_{n}^{\dagger}(x,y)c_{n}(x,y)-\sum_{n=3}^{4}c_{n}^{\dagger}(x,y)c_{n}(x,y)\right],\label{eq:Hm}
\end{equation}
\begin{equation}
H_{y}=-t\sum_{x,y}\left[c_{2}^{\dagger}(x,y)c_{1}(x,y)+c_{4}^{\dagger}(x,y)c_{3}(x,y)+h.c.\right],\label{eq:Hx}
\end{equation}
\begin{equation}
H'_{y}=-t'\sum_{x,y}\left[c_{3}^{\dagger}(x,y)c_{2}(x,y)+c_{1}^{\dagger}(x,y+4)c_{4}(x,y)+h.c.\right],\label{eq:Hx'}
\end{equation}

\begin{align}
H_{x} & =-t_{x}\sum_{x,y}\left[c_{n}^{\dagger}(x+1,y)c_{n}(x,y)e^{i\phi_{n}}\right.\label{eq:Hy'}\\
 & \left.-c_{n}^{\dagger}(x+1,y)c_{n}(x,y)e^{i\phi_{a}}+h.c.\right],\nonumber
\end{align}
with ${\phi}_1=\phi_4=-k_{\varphi}a=\varphi/2$ and $\phi_2=\phi_3 =k_{\varphi}a =-\varphi/2$.

It is convenient to write the Hamiltonian in Fourier space, where it takes
the form $H=\sum_{\mathbf{k}\in BZ1}\psi^{\dagger}(\mathbf{k})h(\mathbf{k})\psi(\mathbf{k}),$
with $\psi=\left( c_{1},  c_{2},  c_{3}, c_{4}\right)^{T}$, and
\begin{equation}
h(\mathbf{k})=\left(\begin{array}{cccc}
m-2t_{x}\cos\left[\left(k_{x}-k_{\varphi}\right)a\right] & -t & 0 & -t'e^{4ik_{y}a}\\
-t & m-2t_{x}\cos\left[\left(k_{x}+k_{\varphi}\right)a\right] & -t' & 0\\
0 & -t' & -m+2t_{x}\cos\left[\left(k_{x}+k_{\varphi}\right)a\right] & -t\\
-t'e^{-4ik_{y}a} & 0 & -t & -m+2t_{x}\cos\left[\left(k_{x}-k_{\varphi}\right)a\right]
\end{array}\right).\label{eq:full Hamiltonian}
\end{equation}
We define a new gauge:
\begin{align}
c_{1}(\mathbf{k})\rightarrow c_{1}(\mathbf{k})e^{-ik_{y}a}, & c_{2}(\mathbf{k})\rightarrow c_{2}(\mathbf{k})e^{-ik_{y}a},\nonumber \\
c_{3}(\mathbf{k})\rightarrow c_{3}(\mathbf{k})e^{ik_{y}a}, & c_{4}(\mathbf{k})\rightarrow c_{4}(\mathbf{k})e^{ik_{y}a},\label{eq:change of gauge}
\end{align}
in terms of which the Hamiltonian takes the simple form
\begin{align}
h(\mathbf{k}) & =\left[m-2t_{x}\cos\left(k_{x}a\right)\cos\left(k_{\varphi}a\right)\right]\sigma_{z}-2t_{x}\tau_{z}\sin\left(k_{x}a\right)\sin\left(k_{\varphi}a\right)\nonumber \\
 & -t\tau_{x}-t'\sigma_{x}\left[\cos\left(2k_{y}a\right)\tau_{x}-\sin\left(2k_{y}a\right)\tau_{y}\right]. \label{eq:h(k)}
\end{align}
In writing the Hamiltonian this way we divide  the $4\times 4$ matrix into four $2\times 2$ blocks, where $\tau_i$ ($i=x,y,z$) are Pauli matrices which operate
within the $2\times2$ blocks separately, while $\sigma_i$ are Pauli
matrices which operate on the outer $2\times2$ matrix of blocks.

\subsection*{Symmetries}

It is easy to check that the Hamiltonian in Eq.~(\ref{eq:h(k)}) has a particle-hole
symmetry,
\begin{equation}
\Lambda h(\mathbf{k})\Lambda=-h^{*}(-\mathbf{k}),\label{eq:particle hole symmetry}
\end{equation}
where $\Lambda=\sigma_{x}\tau_{z}$. The particle-hole symmetry implies
that the bands are symmetric around $E=0$. Therefore, any crossing
of the upper and lower bands must be at $E=0$. Additionally, we have
an inversion symmetry,
\begin{equation}
Ph(\mathbf{k})P=h(-\mathbf{k}),\label{eq:parity symmetry}
\end{equation}
where $P=\tau_{x}$. The crossings therefore must either be at the
high symmetry points, satisfying $-\mathbf{k=k+G}$,
or come in pairs.

\subsection*{Crossing points}

For convenience, we write the Bloch Hamiltonian, Eq.~(\ref{eq:h(k)}),
in the form

\begin{equation}
h=A\sigma_{z}+B\tau_{z}+C\tau_{x}+D\sigma_{x}\tau_{x}+E\sigma_{x}\tau_{y}.\label{eq:h in terms of ABC}
\end{equation}
In terms of these, the spectrum is
\begin{equation}
E=\pm\sqrt{\left(A^{2}+B^{2}+C^{2}+D^{2}+E^{2}\right)\pm2\sqrt{A^{2}B^{2}+A^{2}C^{2}+C^{2}D^{2}}}.\label{eq:spectrum}
\end{equation}
Using this explicit form, it is now easy to find
that the upper and lower bands cross each other at the high symmetry points when
\begin{equation}
m=\pm2t_{x}\cos\left(k_{\varphi}a\right)\pm\sqrt{t^{2}-t'{}^{2}}.\label{eq:crossing points11}
\end{equation}
These crossing points signal a topological phase transition between
a topologically trivial phase and a phase with a non-zero Chern number. The resulting phase diagram, showing the Chern number as a function of $m$ and $\varphi$ is depicted in Fig. \ref{fig:phase}.
\begin{figure}

\includegraphics[scale=0.7]{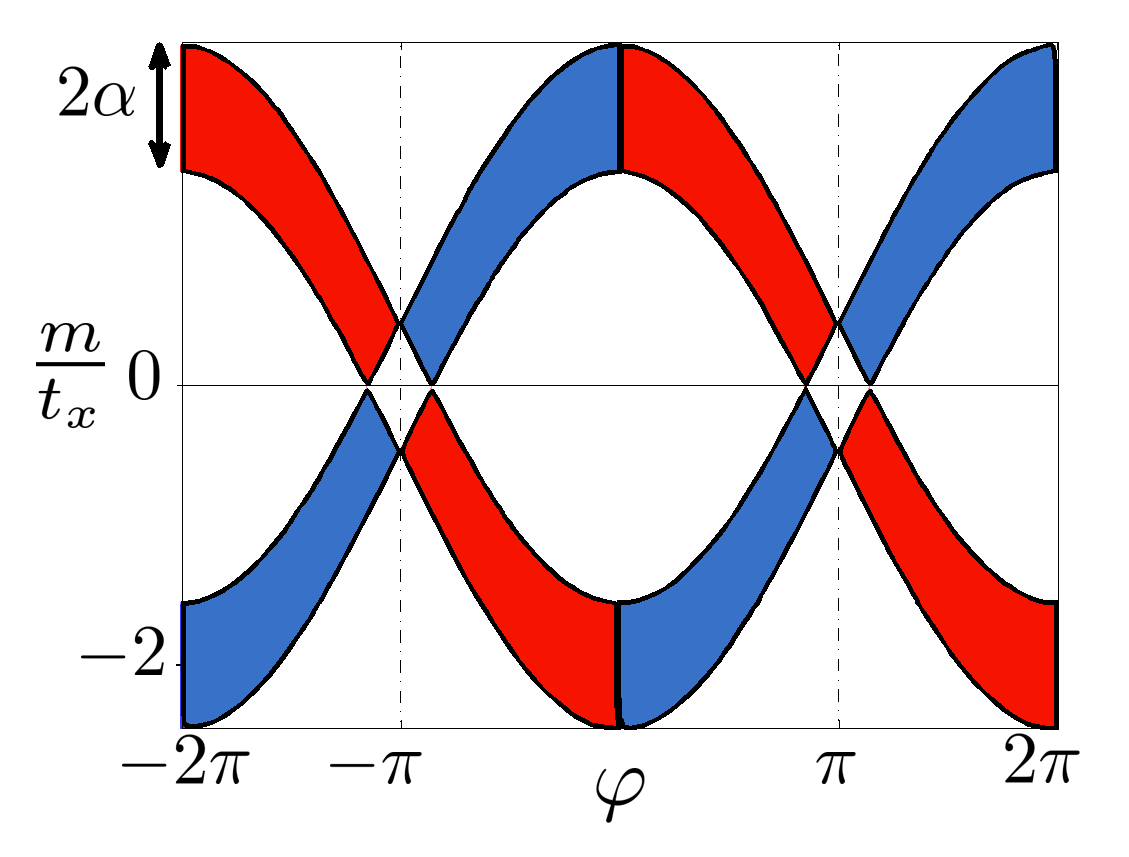}\caption{\label{fig:phase}The phase diagram showing the Chern number as a function of $m$ and $\varphi$. Red regions have $C=1$,
blue regions have $C=-1$, and white regions are topologically trivial
regions with $C=0$. The figure was generated with the parameters $t/t_{x}=0.5,t^\prime /t_x=t^{\prime\prime} /t_x=0.2$}

\end{figure}

There are additional crossings at the points $k_{\varphi}a=0,\pi$, whose locations in $k$-space depend on $m$.
Since these crossings are not at the high symmetry points, they must come in pairs.

If we add a small non-zero chemical potential $\mu$, the bands
cross the chemical potential when
\begin{equation}
m=\pm2t_{x}\cos\left(k_{\varphi}a\right)\pm\sqrt{\left(t\pm\mu\right)^{2}-t'{}^{2}}.\label{eq:chemical potential}
\end{equation}
The topologically trivial and non-trivial insulating phases are now
separated by metallic phases. The transitions between the metallic
and insulating phases are described by Eq.~(\ref{eq:chemical potential}).

In the same way, metallic regions appear near $k_{\varphi}a=0,\pi$.

\section*{Mapping to the self-dual Sine-Gordon model}
In terms of the bosonized $\eta$ fields, the scattering terms presented in the main text are:
\begin{equation}
  v \left[ \cos\left(\eta^L_{2}-\eta^R_{1}\right) +  \cos\left(\eta^R_{3}-\eta^L_{4}\right)
+\left\{ 2\leftrightarrow 4\right\} \right]\label{eq:tunneling between edge states}.
\end{equation}

Defining $\left(
\phi_{c}, \theta_{c}, \phi_{s},\theta_{s}\right)^{T}=U \left(
\eta^R_{1},  \eta^L_{2},  \eta^R_{3}, \eta^L_{4} \right)^T$, with
$$U=\frac{1}{\sqrt{24\pi}}\left(\begin{array}{cccc}
1 & 1 & 1 & 1\\
-1 & 1 & -1 & 1\\
1 & -1 & -1 & 1\\
-1 & -1 & 1 & 1
\end{array}\right),\label{eq:U}$$
we can write the Hamiltonian in the convenient form
\begin{align}
  \int  & \left[ \frac{1}{2}\sum_{a=c,s}\left(\left(\partial_{x}\phi_{a}\right)^{2}+\left(\partial_{x}\theta_{a}\right)^{2}\right)+2\cos\left(\sqrt{6\pi}\theta_{c}\right)\times\nonumber \right. \\
 & \left.  \phantom{\frac{1}{2}} v\left[\cos\left(\sqrt{6\pi }\theta_{s}\right)+\cos\left(\sqrt{6\pi }\phi_{s}\right)\right] \right] dx .\label{eq:hamiltonian density}
\end{align}
If we manage to pin the field $\theta_{c}$, the term $\cos\left(\sqrt{6\pi}\theta_{c}\right)$ can be regarded as a constant, and the resulting Hamiltonian describes the $\beta^{2}=6\pi$ self-dual Sine-Gordon model. We suggest two ways to pin $\theta_{c}$:\\

1. By increasing the bare value of $v$, we can make the operators which multiply $v$ relevant (since the phase diagram is expected to be similar to the Kosterlitz-Thouless-Berezinskii  phase diagram). In this case, the term $\cos\left(\sqrt{6\pi}\theta_{c}\right)$ appearing in Eq.~(\ref{eq:hamiltonian density}) is spontaneously pinned to a minimum, and can be treated as a constant. \\
2. By utilizing the electron-electron interactions, such that the corresponding term is made relevant in the weak coupling limit. 
As an example, we model the Coulomb interactions by
$$H_{int}=v\int dx\rho(x)^{2}, \label{eq:interactions}$$
where $\rho(x)=\frac{\partial_{x}\theta_{c}}{\pi}$ is the total charge density.
In this case the interactions act only on the charge sector, and the kinetic term of the charge sector can be written in the form:
\begin{align}
  \int  & \frac{v^*}{2}\left(K_c\left(\partial_{x}\phi_{c}\right)^{2}+\frac{1}{K_c}\left(\partial_{x}\theta_{c}\right)^{2}\right)dx.\label{eq:modified kinetic term}
\end{align}
When the repulsive interactions become strong enough (i.e. $K_c$ becomes small enough), the term multiplying $v$ becomes relevant.\\
Note that in order to leave a chiral parafermion mode at the edge of the sample, the coupling between different unit cells should be chosen such that it couples the right moving parafermion mode of one unit cell to the left moving parafermion mode of the next unit cell (or vice versa). This was shown to require terms which involve tunneling of quasiparticles [25]. These kind of terms are therefore not allowed between the wires. To overcome this, we need to effectively create a bulk FQHE state between these modes, in which tunneling of quasiparticles is allowed. This can be done if we slightly modify our construction such that each wire contains two flavors (denoted by $a$ and $b$). The spectrum as a function of the momenta $q$'s (corresponding to the composite fields $ \tilde{\psi} $) is shown in Fig.  (\ref{fig:staggered}). We now have effectively 8 wires in each unit cell, out of which we can create an effective bulk FQHE state. Green arrows represent the operators which form the FQHE state, and red arrows represent the operators which create the gapless parafermion modes. The quasiparticle tunneling terms needed to create the $\mathbb{Z}_3$ parafermion state are now allowed by the thin FQHE bulk in each unit cell, and one can construct the desired 2D state.

\begin{figure}

{\includegraphics[scale=0.48]{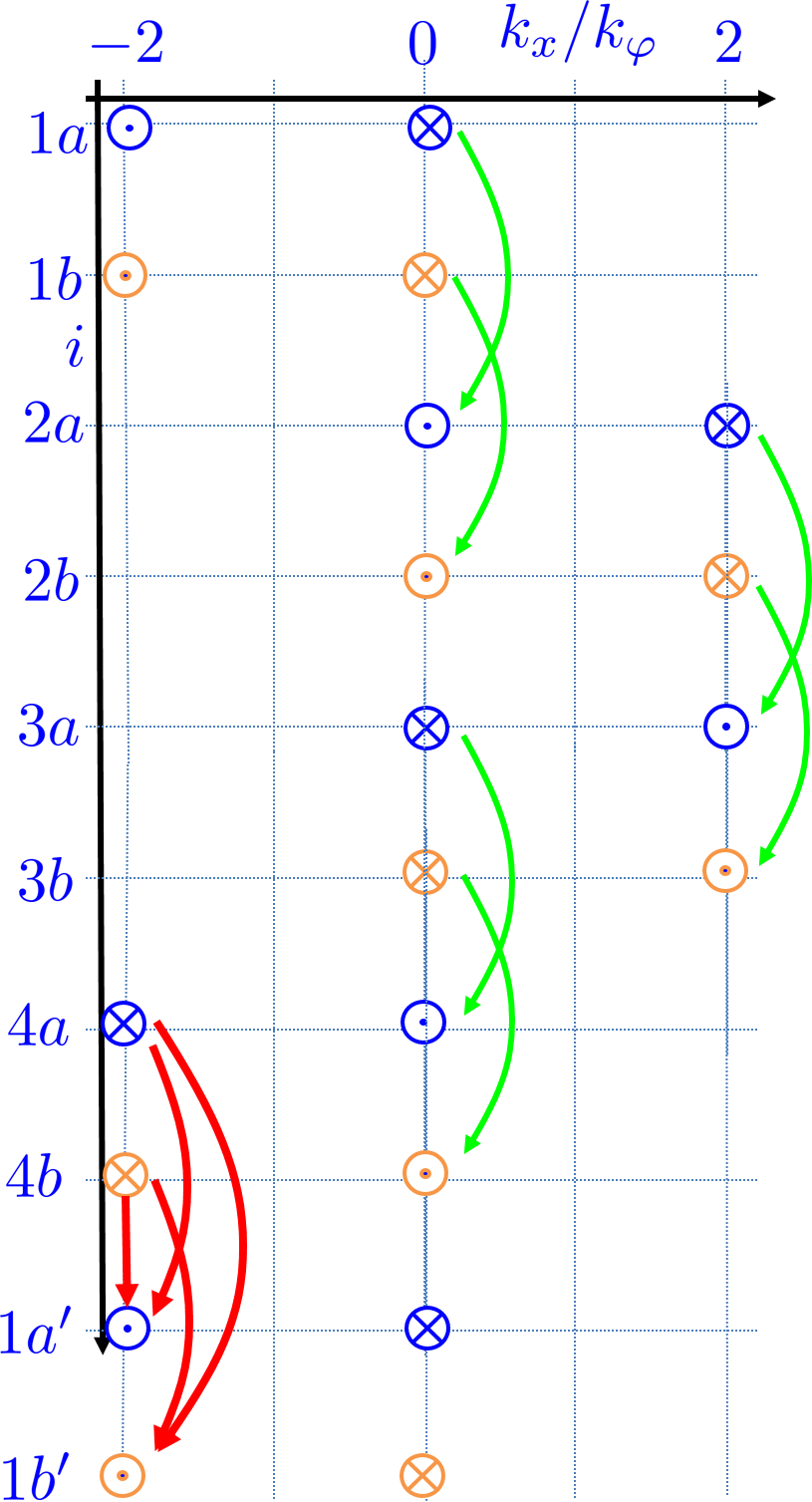}}
\caption{\label{fig:staggered} The diagrammatic representation of the spectrum that corresponds to the case where the system is tuned to $\nu=1/3$ and each wire has two flavors. The momenta shown in the figure are the $q$'s corresponding to the composite fields $ \tilde{\psi} $ defined in the main text. Blue and orange modes represent the flavors $a$ and $b$. Green arrows are terms used to effectively create a bulk FQHE state, and red arrows are used to map the problem to the $\beta ^2 =6\pi$ self-dual Sine-Gordon model in each unit cell. The effective bulk supports the quasiparticle tunneling needed to couple the parafermion modes.}

\end{figure}
This is seen to require further tuning of the couplings [25]. However, due to the topological nature of the phase we describe, deviations from this specific form are possible, as long as the gap in the bulk does not close. In this case, a change in the couplings only affects the width of the edge modes.

\section*{Experimental realization}

Our Q1D construction requires an alternating magnetic field. Experimentally,
this can be realized using a snake-like wire, similar the one
shown in Fig. 3a of the main text.

 In this configuration, assuming that the width of the $p$-$n$ regions is $~\sim 100nm$ and  using the Biot-Savart law we find that the desired current needed to create the states we constructed is of order
\[
I\sim\frac{n\pi\hbar c^{2}}{2e} \approx (100 mA) (100nm) n ,
\]
where $n$ is the one-dimensional density of the conducting band.
Assuming $n\sim (1000 nm)^{-1}$, we get that currents of order $\sim 10mA$
should suffice.
\end{document}